\documentclass[
prl%
 ,twocolumn%
 ,secnumarabic%
,amssymb, amsmath,nobibnotes, showpacs,aps, prl]{revtex4}
\usepackage{epsfig}%
\usepackage{bm}%
\usepackage{graphics}%
\expandafter\ifx\csname package@font\endcsname\relax\else
 \expandafter\expandafter
 \expandafter\usepackage
 \expandafter\expandafter
 \expandafter{\csname package@font\endcsname}%
\fi
\begin{document}
\title{Enhanced  polarization of CMB from thermal gravitational waves}
\author{Kaushik Bhattacharya,
Subhendra Mohanty and Akhilesh Nautiyal} \affiliation{Physical
Research Laboratory, Navrangpura, Ahmedabad - 380 009, India.}

\def\nb{\nabla }
\def\gm{\gamma }
\def\al{\alpha }
\def\op{\oplus }
\def\eps{\epsilon }
\def\dg{\dagger }
\def\pr{\prime }
\def\lm {\lambda }
\def\pl{\parallel}
\def\rar{\rightarrow}
\def\be{\begin{equation}}
\def\lan{\left\langle}
\def\ran{\right\rangle}
\def\jo{J_o}
\def\jpm{J_\pm}
\def\j-{\J_-}
\def\i{\item}
\def\ee{\end{equation}}
\def\wt{\tilde}
\def\be{\begin{equation}}
\def\ee{\end{equation}}
\def\al{\alpha}
\def\bea{\begin{eqnarray}}
\def\eea{\end{eqnarray}}
\def\bearr{\begin{eqnarray}}
\def\bearrs{\begin{eqnarray*}}
\def\eearr{\end{eqnarray}}
\def\eearrs{\end{eqnarray*}}
\def\barr{\begin{array}}
\def\earr{\end{array}}
\def\p{\partial}
\def\vb{\vec{b}}
\def\oc{\Omega}
\def\th{\theta}
\def\sg{\sigma}
\def\o{\omega}
\def\bgt{\bigtriangleup}
\def\bgtd{\bigtriangledown}
\def\non\non{\nonumber}
\def\nn8{\nonumber\\[15pt]}
\def\l{\left}
\def\r{\right}
\def\un{\underline}
\def\ve{\varepsilon}
\def\f{\frac}
\def\ts{\textstyle}
\def\dis{\displaystyle}
\def\jv{\vector{J}}
\def\la{\lambda }

\begin{abstract}
If inflation was preceded by a radiation era then at the time of
inflation there will exist a decoupled thermal distribution of
gravitons. Gravitational waves generated during inflation will be
amplified by the process of stimulated emission into the existing
thermal distribution of gravitons. Consequently the usual zero
temperature scale invariant tensor spectrum is modified by a
temperature dependent factor. This thermal correction factor amplify
the $B$-mode polarization of the CMB by an order of magnitude at
large angles, which may now be in the range of observability of WMAP.
\end{abstract}
 \maketitle

Inflation \cite{inflation}, in addition to solving the horizon and
flatness problems of the standard hot big-bang model, generates a
nearly scale invariant density perturbations, which has been tested
in the observations of the CMBR angular spectrum. One prediction of
inflationary models which has not yet been tested is the existence
of a nearly scale invariant spectrum of gravitational waves
\cite{grwaves,Pilo}. The definitive test of the existence of these
cosmological gravitational waves would be the observation of $B$
mode polarization in the CMB. The recent WMAP three year results
\cite{wmap3} give only an upper bound on the $B$ mode polarization,
 $\frac{(l+1)l}{2 \pi} C^{BB}_{l=(2-6)} < 0.05 (\mu K)^2$.

In this paper we show that if inflation was preceded by a radiation
era then there would be a thermal background of gravitons at the time
of inflation. This thermal distribution of gravitons would have
decoupled close to Plank era. The generation of tensor perturbation
during inflation would be by {\it stimulated emission} into this
existing thermal background of gravitational waves. This process
changes the power spectrum of tensor modes by an extra temperature
dependent factor $\coth(k/2T)$. At large angular scales ($l \leq 30$)
the power spectrum $P_T=A_T\,k^{n_T}$ of gravitational waves generated
during inflation would have a spectral index $n_T= -1-2
\epsilon$, instead of the standard slow roll inflation prediction
$n_T= - 2 \epsilon$ which implies that $C^{BB}_{l=3} \simeq 10 \times
C^{BB}_{l=30}$.  If a thermal enhancement of low $l$ BB modes exists
then it will be observable with WMAP or in the upcoming Planck \cite{Planck}
experiment. In the conclusion, we discuss the implications of
inflationary models from the observations or non-observation of this low
$l$ thermal enhancement.

The tensor perturbations have two independent degrees of freedom which
can be chosen as $h^{+}$ and $h^{\times}$ polarization modes.  To
compute the spectrum of gravitational waves $h({\bf x},\tau)$ during
inflation, we express $h^{(+)}$ and $h^{(\times)}$ in terms of the
creation- annihilation operator as \cite{Gasperini}:
\bea
h^{(i)}({\bf x}, \tau)&=&\frac{\sqrt{16 \pi}}{a(\tau) M_p}\int
\frac{d^3 k} {(2\pi)^{3/2}} [a_{\bf k} \, f_k(\tau) \nonumber\\
&+&a^{\dagger}_{-{\bf k}}\,f_{k}^*(\tau) ]\,e^{i{\bf k} \cdot {\bf x}}
\nonumber\\
&\equiv& \int \frac{d^3 k}{(2\pi)^{3/2}}{h}_{\bf
k}(\tau)\,e^{i{\bf k} \cdot {\bf x}} \,,\nonumber\\
\label{fth}
\eea
where $a(\tau)$ is the scale factor, ${\bf k}$ is the comoving
wavenumber, $k=|{\bf k}|$, and $M_p= 1.22\times 10^{19}{\rm GeV}$ is the
Plank mass and $i=+,\times$. The power spectrum of the tensor
perturbations is defined as:
\be
\langle {h}_{\bf k} {h}_{\bf
k^\prime} \rangle \equiv \frac{2 \pi^2}{k^3} P_{T} \, \delta^3(\bf
k -\bf k^\prime)\,.
\label{power}
\ee
The usual quantization condition
between the fields and their canonical momenta yields $[a_{\bf k}
, a^\dagger_{\bf k^\prime}]= \delta^3({\bf k-k^\prime})$ and the
vacuum satisfies $a_{\bf k}|0\rangle=0$.  If the graviton field
had zero occupation prior to inflation then $[a_{\bf k}
, a^\dagger_{\bf k^\prime}]= \delta^3({\bf k-k^\prime})$ and the
vacuum satisfies $a_{\bf k}|0\rangle=0$.  If the graviton field
had zero occupation prior to inflation then $ \langle a_{\bf
k}^\dagger a_{\bf k}\rangle=0$ and we would obtain a correlation
function $\sim |f_k(\tau)|^2$. However if the graviton field was
in thermal equilibrium at some earlier epoch it will retain its
thermal distribution even after decoupling from the other
radiation fields and its occupation number will be given by:
\be
\langle a_{\bf k}^\dagger a_{{\bf k}'}\rangle =\left(
\frac{1}{e^{k /T} -1}\right)\, \delta^3({\bf k}-{\bf k}')\, .
\label{thermal}
\ee
Using Eq.~(\ref{fth}) and Eq.~(\ref{thermal}) it can be seen that:
\bea
\langle {
h}_{\bf k} {h}_{\bf k^\prime} \rangle
&=&\frac{16\pi|f_k(\tau)|^2}{a^2(\tau)M_p^2} \left(1 +
\frac{2}{e^{\frac{k}{T}}-1}\right)\, \delta^3(\bf k -\bf k^\prime)\,,
\nonumber\\
&=& \frac{16\pi|f_k(\tau)|^2}{a^2(\tau)M_p^2}\,\coth\left[\frac{k}{2
T}\right]\, \delta^3(\bf
k -\bf k^\prime)\,.
\label{R2}
\eea
From the defining relation, Eq.~(\ref{power}), for the tensor power
spectrum and Eq.~(\ref{R2}) we find that the power spectrum for the thermal
inflatons can be expressed in terms of the mode functions $f_k(\tau)$
as:
\be
P_{T}(k)=\frac{8 k^3}{\pi M_p^2}\frac{|f_{k}|^2}{a^2(\tau)}
\,\coth\left[\frac{k}{2
T}\right]\,.
\label{power2}
\ee
The mode functions $f_k(\tau)$ obey the minimally coupled
Klein-Gordon equation:
\be 
f_k^{\prime \prime} + \left(k^2 - \frac{a^{\prime
\prime}}{a}\right)f_k=0\,.  
\ee
In a quasi De Sitter universe during inflation, conformal time $\tau$
($d\tau\equiv dt/a$) and the scale factor during inflation $a(\tau)$
are related by $a(\tau) = -1/H\tau(1- \epsilon)$ where $\epsilon
=\frac{M_{p}^2}{16\pi}\left(\frac{V'}{V} \right)^2$ and $V$ is the
potential for the inflaton field.

For constant $\epsilon$ the mode functions $f_k(\tau)$ obey the
minimally coupled Klein-Gordon equation \cite{Riotto:2002yw},
\be
f_k^{\prime \prime} + \left[k^2 - \frac{1}{\tau^2}\left(\nu^2
- \frac14\right) \right]f_k=0\,,
\label{f2}
\ee
where $k=|{\bf k}|$ and, for small $\epsilon$ and $\delta$,
$\nu=\frac32 + \epsilon$ . Eq.~(\ref{f2}) has the general solution
given by,
 \be f_k(
\tau)=\sqrt{-\tau}\left[c_1(k)\,H^{(1)}_\nu(-k\tau)+
c_2(k)\,H^{(2)}_\nu(-k\tau)\right]\,. \label{sol1} \ee
When the modes are well within the horizon they can be
approximated by flat space-time solutions ${f_k}^0(\tau) =
\frac{1}{\sqrt{2 k}}e^{-i k \tau}\,, (k \gg  a H)$.
Matching the general solution in Eq.~(\ref{sol1}) with the
solution in the high frequency (flat space-time) limit gives the
value of the constants of integration $
c_1(k)=\frac{\sqrt{\pi}}{2}e^{i(\nu +
\frac12)\frac{\pi}{2}}\,~~~{\rm and}~~~~c_2(k) = 0\,$.
Eq.~(\ref{sol1}) then implies that for $-k\tau \gg 1$ or $k \ll
aH$,
\be f_k(\tau)=e^{i(\nu - \frac12)\frac{\pi}{2}}2^{\nu-\frac32}
\frac{\Gamma(\nu)}{\Gamma(\frac32)}\frac{1}{\sqrt{2k}}(-k\tau)^{\frac12-\nu}\,.
\label{superhorizon}
\ee
Substituting the solution as given in Eq.~(\ref{superhorizon}) for the
super-horizon modes ($k\ll aH$) in Eq.~(\ref{power2}) for
the tensor power spectrum, we obtain:
\be P_{T}(k) =\frac{16\pi}{M_P^2}\left(\frac
{H}{2\pi}\right)^2\,\left(\frac{k}{aH}\right)^{n_T}
\,\coth\left[\frac{k}{2 T}\right]\,,
\label{PR2}\\
\ee
with $n_T=3-2\nu=-2\epsilon$. We can now rewrite the power
spectrum as,
\be
P_{T}(k) = A_T(k_0)~\left(\frac{k}{k_0}\right)^{n_T}~
\coth\left[\frac{k}{2 T}\right]\,,
\label{pR}
\ee
where $k_0$ is referred to as the pivot point and $A(k_0)$ is the
normalization constant, $A_T(k_0)=\frac{16\pi}{M_P^2}
\left(\frac{H_{k_0}}{2\pi}\right)^2$ where $H_{k_0}$ is the Hubble parameter
evaluated when $aH=k_0$ during inflation.

The angular power spectrum of the $BB$ polarization modes
generated by the gravitational waves is given by \cite{Seljak:1996gy},
\bea
&C_{l}^{BB}&=(4 \pi)^2 \int dk k^2 P_T(k) \nonumber\\
&\times& \left|\int d\eta g(\eta) h_k(\eta)[2j_l^{\prime
\prime}(x)+\frac{4 j_l(x)}{x}]\right|^2\,,
\label{bl}
\eea
where $g(\eta)=\dot \kappa e^{-\kappa}$ is the visibility function and
$\dot \kappa$ is the differential optical depth for Thomson
scattering. The $EE$ polarization signal gets a contribution from the
tensor perturbations but it is mainly dominated by the scalar
perturbations. So the best signal for gravitational waves is the $BB$
polarization angular spectrum which is generated by the primordial
tensor perturbations only.

The temperature dependent factor becomes important when the ratio
$k/(2 T)$ is less than unity. The co-moving wave-number $k$ and the
co-moving temperature $T$ can be related to the physical parameters
at the time of inflation as follows. Taking the largest measurable
perturbation scale $k_{now}/a_{now} \simeq R_h^{-1}$ (where $R_h
=4000 {\rm Mpc}$ is the size of the present horizon), and assuming
that perturbations of the present horizon scale were just leaving
the inflationary horizon $H^{-1}$ at the beginning of inflation we
see that the temperature at the beginning of inflation $T_i/a_i$
must be,
\be
\frac{k}{2T}=\frac{H a_i}{2T_{i}} < 1\,,
\ee
in order to have a significant effect on the tensor power spectrum.
$T_i/a_i \sim (30 V/g_* \pi^2)^{1/4}$, $V$ being the inflaton
potential which is related to the curvature at the time of inflation,
$H =(8\pi/3)^{1/2} V^{1/2}/M_{P}$ and $g_* \sim 100$ is the effective
number of spin/polarization degrees of freedom of relativistic
particles. Therefore inflation is expected to start at a temperature $
T_i/a_i = 0.24 (H M_{P})^{1/2}$. Actually the gravitons which are
decoupled will have a temperature slightly below the radiation
temperature because of the particles (like the inflaton itself) which
have annihilated into radiation prior to inflation. But as the
effective number of degrees of freedom, $g_* \sim 100$, is large this
difference of temperature is not significant. So for inflation at the
GUT scale, $V^{1/4} \sim 10^{15} {\rm GeV}$, we have $H
\sim 10^{11} {\rm GeV}$ and the temperature at the start of inflation
$T_i/a_i \sim 10^{14 - 15} {\rm GeV} $. So the enhancement of the graviton
power spectrum by the factor $\coth(\frac{k}{2T})=\coth(\frac{H
a_i}{2 T_i})$  could be by as large as a factor of $10^{4-5}$ at low $k$
due to thermal gravitons.

\begin{figure}[h!]
\begin{center}
\epsfig{file=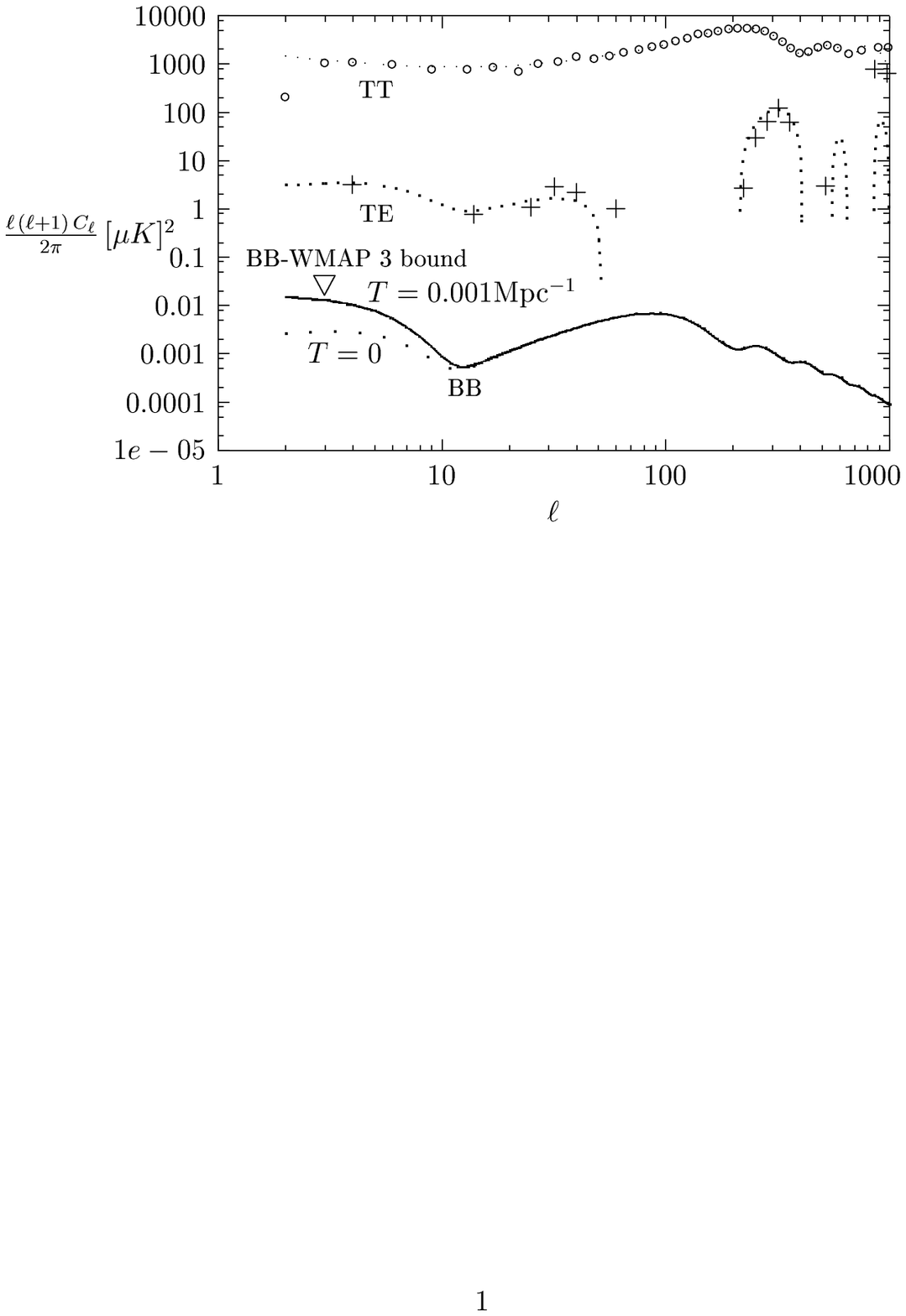,angle=0,width=9cm,height=8cm}
\end{center}
\caption[] {\small\sf The TT, TE and the BB correlations with
thermal graviton spectrum  along with the WMAP three years data
\cite{wmap3}. The plots for TT , TE and BB correspond to co-moving
graviton temperature  $T=.001{\rm Mpc}^{-1}$. For comparison we have
plotted the $BB$ angular correlations at $T=0$. We see that with a
graviton temperature $T=.001{\rm Mpc}^{-1}$ the BB correlations are
amplified at $l<30$.}
\label{f:simul1}
\end{figure}

In Fig.~\ref{f:simul1} we show  the angular correlations of CMBR
temperature and polarization assuming a thermal graviton spectrum (
along with the WMAP three years data \cite{wmap3}). The plots for
TT, TE and BB correspond to co-moving graviton temperature
$T=.001{\rm Mpc}^{-1}$. For comparison we have plotted the $BB$
angular correlations at $T=0$. We see that with a temperature
$T=.001{\rm Mpc}^{-1}$ the BB correlations are amplified at $l<30$.
We see that only the $BB$ correlation is enhanced by the correction
to the tensor power spectrum as expected. The contribution of
tensors to the TT angular spectrum is comparable at low $l$ to the
contribution from the scalars and there exists the possibility that
this large tensor contribution at low $l$ may be detected from the
analysis of the the TT angular spectrum alone.

We have added the unlensed scalar and tensor contributions to generate
the $TT,EE,TE$ and $BB$ correlations. The plots were obtained by
running CMBFAST \cite{CMBFAST}, with the following parameters
${\Omega}_b=0.05$ , ${\Omega}_c=0.25$ and ${\Omega}_v=0.70$. Optical
depth $\tau= 0.08$ and Hubble parameter $h=0.7$. The value of scalar
spectral index $n_s= 0.97$ and the value of tensor spectral index is
taken $n_T=-0.01$. Tensor to scalar ratio is taken to be $r(k_0)= 0.1$
at $k_0=0.002 {\rm Mpc}^{-1}$. The output of the CMBFAST was
normalized to the WMAP values at $k=0.002 {\rm Mpc}^{-1}$ (i.e
$l=30$). For the curves shown in Fig.~\ref{f:simul1} the tensor power
spectra is modified due to thermal effects with $\frac{k}{2T}=500
k$. At $k=0.0002 {\rm Mpc}^{-1}$ , $\coth(500 k)=10$ so there is a
large enhancement of the BB polarization at $l=2-6$, while at
$k_0=0.002{\rm Mpc}^{-1}$ , $\coth(500 k_0) \sim 1.3$ and there is
hardly any enhancement of the BB signal (or in the value of $r(k_0)$
in keeping with the observational constraints from WMAP+SDSS
\cite{scalar-tensor}).  The magnitude of the co-moving graviton
temperature needed to produce this effect is $T/a_{now}=10^{-3} {\rm
Mpc}^{-1}$. This corresponds to a temperature of $T_i/a_i \simeq
4\times R_h^{-1}\times a_{now}/a_i = 4 H$ (where $R_h \sim 4000 {\rm
Mpc}$ is the size of the present horizon). As we have seen inflation
can start as soon as the temperature $T_i/a_i$ falls below $V^{1/4}
\sim 10^4 H$. Consequently a temperature larger than $4 H$ at the
beginning of inflation is not unreasonably high.

In standard inflation models the vacuum fluctuations of the
inflaton field give the density perturbations. The inflaton is
assumed to be decoupled from the radiation at the onset of
inflation, however if there was a radiation era prior to
inflation then the scalar curvature power spectrum is modified by
the same temperature dependent factor \cite{Bhattacharya:2005wn} as the tensor
power spectrum in Eq.~(\ref{PR2}),
\be
 P_{\cal R}(k) =\frac{H^4}{4\pi^2\dot\phi^2}\,\left(\frac{k}{aH}\right)^{n_s-1}
\,\coth\left[\frac{k}{2 T}\right]\,.
\label{PR3}
\ee
The extra temperature dependent term implies that there should be an
up-turn of the $TT$ anisotropy spectrum at low $l$ .  This expected
up-turn in $l(l+1) C_l$ is not seen in the WMAP one-year $TT$ spectrum
\cite{Bhattacharya:2005wn}. This means that there is no significant
number density of background density of inflatons at the time when the
modes which are currently entering our horizon, were exiting the
horizon during inflation. This could happen for two reasons. The
background density of inflatons may have decayed or annihilated into
lighter particles by this time or the inflaton was cooled from the
expected temperature of $0.24 (H M_P)^{1/2}$ to below $H$ by the time
the modes corresponding to our present horizon were leaving the
De-Sitter horizon. This implies that there were an extra $\Delta N=
\ln (0.24 (M_P/H)^{1/2})$ e-foldings (which has the value $\Delta N
\sim 10$ for GUT scale inflation) than what is needed to solve the
horizon problem. In the case of gravitons the first condition
does not apply as they decouple at the Planck scale and if the expected
upturn in the $BB$ mode spectrum is not seen that would imply that
the duration of inflation was longer than what is needed to solve
the horizon problem.

In warm inflation models \cite{warm} where the inflaton is in thermal
equilibrium with the radiation bath and the scalar curvature
perturbations are generated by thermal fluctuations instead of by
quantum fluctuations, there is no $\coth(k/2T)$ correction in the
inflaton power spectrum due to stimulated emission. However this
correction factor will be present in the graviton spectrum since
gravitons are still produced by quantum fluctuations, in warm
inflationary models. The scalar curvature perturbation in warm
inflation is \cite{warm2}:
\bea
P^{(\rm warm)}_{\cal R}= \left(\frac{\pi}{4}\right)^{1/2}
\frac{H^{5/2} \Gamma^{1/2} T_r}{\dot \phi^2}\,,
\label{PRwarm}
\eea
where $\Gamma$ designates the decay width of the inflaton field and
$T_r$ is the temperature of the radiation bath.

There are observational constraints on the tensor scalar ratio defined as:
\be
r(k_0)= \frac{P_T(k_0)}{P_{\cal R}(k_0)}\,.
\ee
From the combination of WMAP three year data \cite{scalar-tensor}
and SDSS large scale structure surveys \cite{sdss} we have the bound
$r(k_0=0.002 {\rm Mpc}^{-1}) < 0.28 (95\% CL)$ where $k_0=0.002 {\rm
Mpc}^{-1}$ corresponds to $l=\tau_0 k_0 \simeq 30 $ with the
distance to the decoupling surface $\tau_0=14,400 {\rm Mpc}$. SDSS
measures galaxy distributions at red-shifts $a \sim 0.1$ and probes
$k$ in the range $0.016 h {\rm Mpc}^{-1} < k <0.11 h {\rm
Mpc}^{-1}$. From the expressions of $P_{\cal R}$ in warm inflation,
Eq.~(\ref{PRwarm}), and $P_T$ we see that the scalar-tensor ratio in
warm inflation models (assuming a nearly scale invariant tensor power
spectrum) has a scale dependence at large angles given
by:
\be r(k)\simeq r(k_0) \frac{\coth[\frac{k}{2 T}]}{\coth[\frac{k_0}{2
T}]}\simeq r(k_0)
 \left(\frac{k_0}{k} \right)\,.
\label{rcond}
\ee
We see that $r(k)$ has a spectral index $n_T \sim -1$ for large
scale perturbations. If we consider $k \sim 0.0002 {\rm Mpc}^{-1}$
which corresponds to $l\sim 3$ then the value of $r(k)=10 r(k_0)$.
So even with $r(k_0)\sim 0.1$ as constrained by galaxy surveys, we
can have $r(k) \simeq 1$ at the quadrupole anisotropy. The $B$
mode polarization at $l=3$ is enhanced from its value at $l=30$ by a
corresponding factor of $10$. This is true as long as the
temperature $T_i/a_i \leq 10^4 H$ which as we have seen in the earlier
discussion is expected if there is a thermal era prior to inflation.

For example taking the inflaton potential to be $V=(1/2)m^2 \phi^2$,
we have the scalar power:
\be P^{(\rm warm)}_{\cal R}(k_0)= 5.3 \frac{\Gamma^{5/2}
\phi_0^{1/2} T_r}{M_p^{5/2} m^{3/2}}\,,
\ee
and the tensor power,
\be P_T(k_0)= \frac{128\pi}{9} \frac{m^2
\phi_0^2}{M_P^4}\coth\left[\frac{k_0}{2T}\right]\,,
\ee
and the scalar-tensor ratio,
\be r(k_0)= 8.413\left(\frac{m^{7/2} \phi_0^{3/2}}{M_P^{3/2}} \right)
\frac{1} {\Gamma^{5/2}T_r} \coth\left[\frac{k_0}{2T}\right]\,,
\ee
where $\phi_0$ is the value of the inflaton field when the scale
$k_0=0.002{\rm Mpc}^{-1}$ was leaving the inflaton horizon.  By
choosing the parameters $m=1.4 \times 10^{12} {\rm GeV}$, $\Gamma=0.5
\times 10^{13} {\rm GeV}$, $T\simeq T_r=0.24 \times 10^{16}{\rm GeV}$,
$\phi_0 \simeq 0.8\times 10^{19}{\rm GeV}$ we have $P_R \simeq 2.3
\times 10^{-9}$ as required by WMAP three year data and $r(k_0)
=0.095$. The value of $r$ is larger at $k=0.0002 {\rm Mpc}^{-1}$ by a
factor of $\sim 10$ and the $B$-modes are magnified at $l=3$ compared
to their value at $l=30$ by a factor 10, also in warm inflation
scenarios.

Direct observation of gravitational waves would nail the last still
unconfirmed prediction of inflation. The amplitude of gravitational
waves gives the Hubble curvature during inflation and would tell us
the value of the inflation potential \cite{knox}. In addition gravitational
waves produced during inflation can have several applications like
leptogenesis by the gravitational spin-coupling to neutrinos
\cite{mohanty} or by a gravitational Chern-Simon coupling of the
lepton number current \cite{peskin}. Observation of the $B$-mode
polarization in the CMB would confirm the existence of primordial
super-horizon gravitational waves. Observationally, the three year
WMAP data only gives an upper bound on $C_l^{BB}$ with $l=(2-6)$
\cite{wmap3}. The error bars on the $C_l^{BB}$ are presently a
factor of five larger than the predictions from standard inflation
theory with scalar tensor ratio as large as $0.3$, which is close to
the observational upper bound $r_{0.002}< 0.28 (95\% CL)$. In this
paper we show that due to thermal gravitons, the $C_l^{BB}$ at low
$l\simeq (2-6)$ could be larger by a factor of $10$ compared to what
would be expected from the observational constraint on $r$ and could
be within the range of observability of WMAP. The upcoming Planck
experiment \cite{Planck} will measure $C_{l=(1-10)}^{BB}$ at the level
of $ 10^{-4} (\mu K)^2$ . Ground based polarization experiments
\cite{Verde} like QUaD, QUIET, Clover and PolarBear measure
anisotropies at small angular scales only (at $l>100$ where thermal
effects discussed in this paper are negligible) and can observe
$C_l^{BB}$ at the level $ 10^{-2} (\mu K)^2$ . These experiments can
probe $r$ in the range $ 0.05-0.1$ independent of thermal effects. A
combination of data from WMAP/Plank at large angles and ground based
polarization experiments at small angles will therefore either observe
or definitely rule out the thermal enhancement effect.

If WMAP or Planck  rule out a spectral index of $n_T \sim
-1$ at low $l$, which is the prediction from thermal gravitons, then
 for the standard inflationary models it would mean that the duration of inflation has
to be longer by $\Delta N= \ln (0.24 (M_P/H)^{1/2})$ e-foldings than
what is needed to solve the horizon problem. Warm inflation
models\cite{warm,warm2} cannot evade this constraints by
supercooling during inflation. If $B$-modes are observed and the
tensor spectral index at low $l$ is not close to $-1$, then warm
inflation models can be ruled out.


\end{document}